\documentclass[aps,prb,twocolumn,showpacs,groupedaddress]{revtex4}

\usepackage{graphicx}
\usepackage{amssymb}

\begin{document}
\setlength{\unitlength}{1cm}

\title{On ground states of interacting Composite Fermions with spin at half
  filling.}

\author{Matteo Merlo$^{1}$, Nicodemo Magnoli$^{2}$, Maura Sassetti$^{1}$
and Bernhard Kramer$^{3}$}
\affiliation{
  $^{1}$Dipartimento di Fisica, INFM-LAMIA, 
  Universit\`{a} di Genova, Via
  Dodecaneso 33, I-16146 Genova, Italy \\
  $^{2}$Dipartimento di Fisica, INFN, 
  Universit\`{a} di Genova, Via
  Dodecaneso 33, I-16146 Genova, Italy \\
  $^{3}$I. Institut f\"ur Theoretische Physik,
  Universit\"at Hamburg,
  Jungiusstra\ss{}e 9, D-20355 Hamburg, Germany }

\date{\today}
\begin{abstract}
 The effects of interactions in a 2D electron system in a strong magnetic
  field of two  degenerate Landau levels with opposite spins and at
  filling factors $1/2$ are studied. Using the Chern-Simons gauge
  transformation, the system is mapped to Composite Fermions. The fluctuations
  of the gauge field induce an effective interaction between the Composite
  Fermions which can be attractive in both the particle-particle  and in the particle-hole channel. 
  As a consequence, a spin-singlet (s-wave) ground state of Composite Fermions can exist with a
  finite pair-breaking energy gap for particle-particle or particle-hole
  pairs. The competition between these two possible ground states is
  discussed. For long-range Coulomb interaction the particle-particle state is
  favored if the interaction strength is small.  With increasing interaction
  strength there is a crossover towards the particle-hole state. If the
  interaction is short range, only the particle-particle state is possible.

\end{abstract}

\pacs{71.10.Pm; 73.43.Cd; 73.43.Nq}

\maketitle

%*****************************************************************************
%                                                             INTRODUCTION
%*****************************************************************************

\section{Introduction}
The Composite Fermion (CF) model for the fractional quantum Hall effect (FQHE) has 
been very successful in describing in a simple and intuitive way the basic filling 
factors at which this complex collective phenomenon occurs. \cite{Jain89} One route 
to CFs is the Chern-Simons (CS) gauge transformation which maps a system of 
interacting electrons in a Landau level (LL) at an even-denominator fractional 
filling into a weakly interacting Fermi liquid of CFs.\cite{SternHalp95,LF91,HLR93} 
This is achieved by formally attaching an even number of flux quanta to each 
electron.  On the average, the CFs do not see the external magnetic field but a 
smaller effective one, which vanishes in mean field approximation at the 
even-denominator fractional filling considered. The Coulomb interaction between the 
electrons is in this model incorporated into a finite effective mass.\cite{Books} 
The incompressible states responsible for the FQHE of the electrons can then be 
described in terms of integer quantum Hall states of the CFs. Experimental support 
for the model comes from measurements near filling factor one half.\cite{willett} 
Theoretical expectations concerning the properties of the CFs \cite{HLR93} have been 
confirmed by surface-acoustic wave\cite{saw} and transport experiments in 
periodically modulated structures\cite{anti} and from cyclotron resonance.\cite{k02} 

The results obtained until now suggest that constructing compound quasi particle 
states made of charges and fluxes in such a way that the repulsive interaction is 
minimized is a very efficient way of dealing with strongly interacting many particle 
systems. Better understanding of such states may be of great importance beyond 
explaining the fractional quantum Hall effect. High-$T_{\rm 
c}$-superconductivity\cite{bm86}, the unique properties of heavy fermion 
systems\cite{gs91}, and the recently discovered metal-insulator transition in low 
density two dimensional electron systems\cite{ks03} can be suspected to be candidate 
systems where the concept of compound charge-flux quasi particles may eventually 
turn out to be crucial for understanding the underlying correlations. Thus, one is 
led to conclude that studying the physics of charge-flux states is an important 
subject of research in its own right.

At high magnetic field one often can safely assume that the spins are frozen
such that the quantum Hall states are spin polarized. However, due to the
small value of the electron $g$-factor in GaAs ($\approx -0.4$), this
assumption is not always valid, especially for the smaller magnetic field
strengths sufficient to enter the region of the FQHE in the lowest Landau level
for samples with low electron density. It has been experimentally established
that, depending on the filling factor, FQHE states may be
unpolarized\cite{cetal89,eetal89,eetal90,eetal92} ($\nu=4/3, 8/5, 10/7, 2/3$)
or partially polarized ($\nu=3/5, 7/5$).\cite{cetal89,eetal92} There are also
crossovers between different polarizations  when changing the Zeeman splitting 
by tilting the magnetic field, or when reducing the electron density. The
spin polarization of several FQHE states has been optically determined
\cite{Kukushkin} at fixed filling factors as a function of the ratio between
Zeeman and Coulomb energies,
\begin{equation}
  \label{eq:0}
  \xi = \frac{E_{\rm Z}}{E_{\rm C}}\, 
\end{equation}
where 
\begin{equation}
  \label{eq:ec}
  E_{\rm C}=\frac{e^{2}}{\epsilon l_{B}}
\end{equation} 
and $\epsilon$ and $l_{B}=\sqrt{\hbar c/eB}$ are respectively the dielectric 
constant and  the magnetic length.

 Crossovers between differently spin polarized ground states for the same FQHE 
filling factor have been detected. The spin polarization remains constant within 
large intervals of $\xi$. Near certain critical values $\xi_{\rm cr}$, the system 
undergoes a transition between differently spin-polarized CF states. A simple model 
of non-interacting CFs with spin with an effective mass that scales as the Coulomb 
interaction, i.e. $m^{*}\propto\sqrt B$, can explain the experimental data. The 
broad plateaus of constant spin polarization are due to the occupation of a fixed 
number of spin split LL of the CFs (CFLL). The crossovers occur when intersections 
of CFLLs with opposite spins coincide with the chemical potential.\cite{annalen}

The optically determined spin polarizations, when extrapolated to zero temperature, 
show additional plateaus for flux densities near the centers of the crossovers. The 
corresponding polarizations are almost exactly intermediate between those in the 
neighboring broad plateaus within the experimental uncertainties. This indicates 
additional physics beyond the non-interacting CF model. The intermediate plateaus 
can be interpreted as the signature of new collective states since one can expect 
that if two CFLLs are degenerate, interactions between CFs become very important and 
cannot be treated perturbatively. In these optical experiments, the CFLL have been 
tuned to degeneracy by using the magnetic field dependence of the effective mass of 
the CFs. Intermediate plateaus have also been observed with NMR where $\xi$ was 
changed by tilting the magnetic field.\cite{fetal01}
 
Recent experimental studies of the FQHE in GaAs/AlGaAs samples of densities $\approx 
10^{11}$\,cm$^{-2}$ revealed strong FQHE-structures at filling factors $\nu=4/11$ 
and $5/13$ and weaker structures at $6/17$, $4/13$, $5/17$ and $7/11$.\cite{petal03} 
The feature at $4/11$ is independent of an in-plane component of the magnetic field 
and is expected to be spin polarized. These new FQHE states cannot be explained 
within standard sequences of IQHE of CFs. It seems that rather they are signature of 
a FQHE of CFs. This could imply that interaction between CFs can be expected to be 
strong.

One may summarize the above observations by noting that on the one hand spin
is an important ingredient of the physics of composite charge-flux quasi
particles that must not be neglected, and on the other hand that the
interactions between the quasi particles may lead to qualitatively new
collective quantum states. Better understanding of the latter, especially in
the presence of spin, seems imperative not only for explaining the rich
phenomena of the physics of the FQHE\cite{m03} but could also lead eventually
to new insights into the physics of low dimensional many body systems.

Without using the CF model, several possibilities for the states that could form 
under the above conditions have been discussed.\cite{Murthy00,a01} However, in order 
to systematically understand interaction-induced and spin polarization properties of 
the FQHE states, the CF model can be expected to be useful.\cite{pj98} The first 
step is to generalize the CS-transformation to include the electron spin. With this
supplementary degree of freedom, useful analogies can be drawn with bilayer systems 
of spinless fermions, where the electrons carry a layer - instead of a spin - index.
This generalization to 2-component systems of CFs with index $s=\uparrow,\downarrow$ (or $1,2$)
can be achieved with models in which a doublet of  Chern-Simons gauge fields is introduced\cite{lopfrad01}; its 
Lagrangian  contains a $2\times2$ matrix $\Theta_{ss'}$
\begin{eqnarray}
  \label{eq:kdd}
\Theta=\left(
  \begin{array}{cc}
\theta_{1}&\theta_{2}\\
\theta_{2}&\theta_{1}
  \end{array}\right)\,,
\end{eqnarray}
 which controls the attachment of flux quanta
to the two species of fermions [see (\ref{eq:constraint}) below for details]. 
 In the spin case, with such an approach  
many of the FQHE wave functions proposed hitherto for FQHE systems, with their spin 
polarizations, have been reproduced.\cite{mr96}

 In the bilayer system at total filling factor 1, and such that in each 
layer $\nu=1/2$, it has been argued\cite{ketal01,m02} that for small layer distance
 a spin-polarized p-symmetric pair state can be formed that is equivalent to the 
so-called (1,1,1)-state proposed earlier.\cite{h83} This state consists of pairs of 
interlayer (or \emph{mutual}\,\cite{v02,ye03}) CFs: $\Theta$ is chosen in such a way 
($\theta_1=0,\theta_2=1/2$) that an electron in one layer is attached to two flux 
quanta in the other layer and vice versa.  In this language, the CFs are 
attractively interacting  interlayer dipolar objects, due to the fluxes being 
equivalent to ''holes'' in the electron system; this is believed to be a possible 
mechanism for the interlayer phase coherence recently found in experiments in this 
regime.\cite{sp00} In general, different choices of $\Theta$   that preserve the 
fermionic statistics of the original particles can be exploited to describe the 
system when the layer distance is varied: a diagonal attachment of 2 flux quanta 
($\theta_1=1/2,\theta_2=0$) is thought to describe correctly the intermediate- and 
large- distance regime of bilayers.\cite{ketal01}

In the single layer with spin, generalized CFs have been introduced by a {\em 
non-unitary} Rajaraman-Sondhi\cite{raj96}  instead of the 
CS-transformation.\cite{m00} The effective interaction between them contains the 
repulsive long-range Coulomb part, a contribution due to the gauge field 
fluctuations and a non-Hermitian term that destabilizes the CF states. Neglecting 
the non-Hermitian term, it was found that due to the {\em symmetry of the gauge 
field term} in the electron-electron interaction, which enters here in first order, 
s-wave pairing is not possible in static mean field approximation. By estimating the 
condensation energies it was found that if a pair state at total filling factor 1/2 
was realized it would be a spin polarized p-wave state.
 Due the static mean field character of the approximation used, off-diagonal terms in
the matrix $\Theta$ are needed in this approach to couple the two spins.

In the present paper, we reconsider the effective interaction between CFs with spin. 
Especially, we concentrate on the competition between the formation of 
particle-particle (p-p)  and particle-hole (p-h) pairs in the s-wave channel. We 
consider a spin degenerate lowest Landau level at filling factor unity and assume 
that $N$ electrons are distributed among the available states in such a way that 
exactly half of them have spin $\uparrow$ and the other half have spin $\downarrow$. 
This is equivalent to two degenerate half-filled LL with opposite spins such that 
for each the CS transformation can be applied in order to obtain CFs. We assume  
that only the diagonal parts of the coupling matrix are non-zero,
\begin{eqnarray}
  \label{eq:kdd1}
\Theta=\left(
  \begin{array}{cc}
1/2&0\\
0& 1/2
  \end{array}\right)\,.
\end{eqnarray}
The two subsystems are then transformed to two Fermi seas with $\uparrow$ and
$\downarrow$ coupled by the effective CF interaction. We show that under this
condition, the CS gauge field fluctuations can mediate an attractive interaction 
between CF-particles. In order to obtain this interaction in lowest order, we need to
take into account an RPA-like renormalization of the gauge field fluctuations by the 
coupling to the CFs. The attractive interaction can result in a spin-singlet s-wave 
bound state of pairs of CF particles.\cite{metal02,pss,varenna} Alternatively, 
CF-holes and CF-particles may be bound together, thus forming an excitonic 
spin-singlet state. We consider the competition between the latter exciton-like and 
the former Cooper pair-like  pairings in the CF system. We determine the 
corresponding pair breaking energy gaps, and the ground state energies. We discuss 
the stability of the different phases. For Coulomb interaction between the electrons 
we find that when the interaction strength measured by the Coulomb energy  $E_{\rm 
C}$ is small the Cooper pair-like phase is more stable. When $E_{\rm C}$ is large 
compared with the chemical potential, the exciton-like phase is more stable. For 
short range interaction, the Cooper pair-like phase has always the lowest energy. We 
conjecture that the paired singlet phases are very likely to approximate the ground 
state of the interacting electron system even if the lowest LL is only close to spin 
degeneracy. Then, at zero temperature, the energetically lower LL with, say, spin 
$\uparrow$ will be occupied.  However, if the gain in the ground state energy by 
forming a pair exceeds the cost in energy for occupying a state in the LL with spin 
$\downarrow$, pairs will be formed and the system will condense into the 
spin-singlet ground state.

The above model does not exactly match the situations in the aforementioned
optical experiments where CFLLs with opposite spins corresponding to different
Landau quantum numbers coincide. However, the second generation CFs can provide
the scheme for understanding the additional plateaus at intermediate spin
polarizations.\cite{metal02,annalen} In any case, we feel that the effect of
the residual interactions between CFs and whether or not they can give rise to
new features is an interesting problem in its own right and deserves intense
studies.

The paper is organized as follows. In section \ref{section:effint} the methods
used to determine the effective interaction and the ground state are
described. In section \ref{longrange}, the particle-particle (p-p) and the
particle-hole (p-h) ground-state energies are calculated for 
Coulomb interaction. In section \ref{shortrange} the results are provided for
short-range interaction. The phase-diagrams for the ground states are derived.
Discussion of the results and final remarks will conclude the
paper.

%*****************************************************************************
%                            EFFECTIVE INTERACTION AND GROUND STATE PROPERTIES
%*****************************************************************************

\section{Effective interaction and ground state properties}
\label{section:effint}

We consider two half-filled LL with opposite spins at the same energy. The
CS-transformation is used to construct two 2D Fermi seas of CFs with spin
and a Fermi wave number $k_{\rm F}=\sqrt{2\pi \rho}$ ($\rho$ total average
electron number density).\cite{HLR93} An effective interaction between the CFs
can be obtained from the Lagrangian density of the two coupled Fermi systems of
charge $e$ (units $\hbar=c=1$),
\begin{equation}
{\cal L} ({\bf r},t) = {\cal L}_{\rm F} ({\bf r},t) + {\cal
L}_{{\rm CS}} ({\bf r},t) + {\cal L}_{{\rm I}} ({\bf r},t)
\label{eq:1}
\end{equation}
with the kinetic energy of the Fermions
\begin{eqnarray}
{\cal L}_{\rm F}({\bf r},t)&=&\sum_{s=\uparrow,\downarrow}
\psi^{\dagger}_{s}({\bf r},t)
\Big\{i\partial_{t}+\mu+e a_0^{s}({\bf r},t)\Big.\nonumber \\
-\frac{1}{2m^{}}\Big[i\nabla&+&e \Big({\bf A}({\bf r})-{\bf a}^{s}({\bf 
r},t)\Big)\Big]^2 \Big\}\psi_{s}({\bf r},t) \label{eq:2}
\end{eqnarray}
($m^{}$ effective mass, $\mu$ chemical potential), the CS term
\begin{equation} \label{eq:lcs}
{\cal L}_{{\rm CS}}({\bf
r},t)=-\frac{e}{\phi_{0}}
\sum_{s,s'} \Theta_{ss'} \,a_0^{s}({\bf r},t)\, {\bf
\hat{z}}\cdot\nabla\times{\bf a}^{s'} ({\bf r},t)
\end{equation}
($\phi_{0}=hc/e$ flux quantum, ${\bf \hat{z}}$ unit vector perpendicular to
the 2D plane), and the contribution of the electron-electron interaction
\begin{equation} \label{eq:coulomb}
{\cal L}_{{\rm I}}({\bf r},t)=-{\frac{1}{2}}\sum_{s,s'} \int d^2
r' \rho_{s}({\bf r},t) V({\bf r}-{\bf r}') \rho_{s'} ({\bf r}',t).
\end{equation}
Here, $\rho_{s}({\bf r},t)\equiv \psi^{\dagger}_{s}({\bf r},t) \psi_{s}({\bf
  r},t)$ is the density of the Fermions with spin orientation $s$, ${\bf A}$
the vector potential of the external magnetic field, $(a_{0},{\bf a})$ the CS
gauge field, and $V({\bf r})$ the electron-electron interaction potential. The
attachment of flux quanta $\phi_{0}$ to each Fermion is
achieved by the Chern-Simons term ${\cal L}_{{\rm CS}}$ as can be seen by
minimizing the action with respect to the $a_{0}^{s}$-gauge field. This gives
the constraint 
\begin{equation}   \label{eq:constraint} 
\sum_{s'}\Theta_{ss'} {\bf
\hat{z}}\cdot\nabla\times{\bf a}^{s'} ({\bf r},t) = \phi_0 \rho_s({\bf r},t).
\end{equation}
The flux attachment for the two species of Fermions is in this paper performed
independently. This corresponds to assuming the coupling matrix to be
diagonal: 
\begin{eqnarray}
  \label{eq:kddiag}
\Theta=\left(
  \begin{array}{cc}
\theta_{1}& 0 \\
 0 &\theta_{1}
  \end{array}\right).
\end{eqnarray}
We assume $\theta_1=1/2$, such that the mean fictitious magnetic field 
cancels the external one at half filling, $\nu\equiv \rho\phi_{0}/2B=1/2$. We
use the transverse gauge, $\nabla\cdot{\bf a}^{s}=0$. The Bosonic variables
associated with the gauge field fluctuations are the {\em transverse}
components of their Fourier transforms, $a_{1}^{s}({\bf q},\omega)\equiv {\bf
  \hat{z} } \cdot\hat{\bf q}\times[{\bf a}^{s}({\bf q},\omega)-\langle{\bf
  a}^{s}({\bf q},\omega)\rangle]$.  From the terms linear in the charge $e$
and the momentum $-i\nabla$, one can extract the form of the vertices
connecting two Fermions with one gauge field fluctuation operator
$a_{\mu}^{s}({\bf q},\omega)$ ($\mu=0,1$)
\begin{equation}
\label{eq:vertex}
v^{s}_{\mu}({\bf k},{\bf q})=\biggl(
\begin{array}{c}
e\\
\frac{e}{m}\,{\bf \hat{z}} \cdot\frac{{\bf k}\times{\bf q}}{|{\bf q}|}
\end{array}
\biggr).
\end{equation}
In addition, there is a Fermion-gauge field coupling term quadratic in the
fluctuations $w_{\mu\nu}^s=-\delta_{\mu,1} \delta_{\nu,1}\,e^2/2m^{}$.

 By 
introducing the mean gauge field into ${\cal L}_{\rm F}$ the external field ${\bf 
A}$ is canceled. By Fourier transforming we find for the action $S=\int d{\bf r} dt 
{\cal L}({\bf r},t)$
\begin{widetext}
\begin{eqnarray}
\label{eq:action}
S &=& \sum_s \frac{1}{(2\pi)^3} \int{\rm d}{\bf k} {\rm d}\omega
\psi^\dagger_s({\bf k}, \omega) (G^{0}_s)^{-1}({\bf k},\omega)
\psi_s({\bf k}, \omega) +  \sum_{\alpha, \mu, \nu}
\frac{1}{2(2\pi)^3} \int {\rm d}{\bf q} {\rm d}\Omega
a_\mu^\alpha({\bf q},\Omega) (^0\!D^{\alpha})^{-1}_{\mu\nu}({\bf
q},\Omega) a_\nu^{\alpha\dagger}({\bf q},\Omega) + \nonumber\\
&+& \sum_{s,\mu} \frac{1}{(2\pi)^6} \int {\rm d}{\bf k} {\rm
d}\omega {\rm d}{\bf q}{\rm d}\Omega \psi^\dagger_s({\bf k+q},
\omega+\Omega) \psi_s({\bf k}, \omega) a_\mu^s({\bf q},\Omega)
v_\mu^s({\bf k,q})
\end{eqnarray}
\end{widetext}
with the Green functions of the free Fermions
\begin{equation}
  \label{eq:gffermion}
G^{0}_{s}({\bf
  k},\omega)=\frac{1}{\omega -k^{2}/2m^{}+\mu+i\delta{\rm sgn}\,\omega}\,.    
\end{equation}
The second term in (\ref{eq:action}) consists of ${\cal L}_{\rm CS}+{\cal
  L}_{\rm I}$ and describes the free gauge field. It is obtained by inserting
the constraint (\ref{eq:constraint}) between the charge density and the gauge
field into (\ref{eq:coulomb}), introducing symmetric and antisymmetric
combinations of the gauge field fluctuations
($\alpha=\pm$)
\begin{equation}
  \label{eq:gaugefields}
a_\mu^\alpha=\frac{a_\mu^\uparrow+\alpha a_\mu^\downarrow}{2}\,,  
\end{equation}
and defining
\begin{equation}
  \label{eq:gaugepropagator}
\! (^{0}\!D^{\alpha})^{-1}_{\mu\nu}({\bf q},\Omega)\!=\!\left(\!\!
  \begin{array}{ccc}
 0 && \frac{ieq} {\phi_0} \\
&&\\
-\frac{ieq}{ \phi_0} & 
&-\frac{e^{2}\rho}{m^{}}
-\frac{q^2 V(q)}{ \phi_0^2}\,\delta_{\alpha,+}\\
  \end{array} \!\!\right)
\end{equation}
with the Fourier transformed  interaction potential $V(q)$.

\subsection{The effective interaction.}
\label{sec:effint}
In the following, it turns out to be convenient to proceed with the finite
temperature Matsubara formalism. Thus, we introduce imaginary time Green
functions ($T_{\tau}$ time ordering operator)
\begin{eqnarray}
{\cal G}_{ss'}({\bf k},\tau)&=&-\langle T_\tau \psi_s({\bf
k},\tau)\psi_{s'}^\dagger({\bf k},0) \rangle  \\
{\cal D}_{\mu\nu}^\alpha({\bf q},\tau)&=&-\langle T_\tau
a_\mu^\alpha({\bf q},\tau) a_\nu^{\alpha\dagger}({\bf q},0)
\rangle\,.
\end{eqnarray}
The effective CF interaction can then be obtained from the coupling terms in ${\cal 
L}_{\rm F}({\bf r},t)$. At imaginary time, one gets the kernel of the interaction in 
the frequency domain
\begin{eqnarray}
&V_{\mu\nu}^{s,s'}({\bf k},{\bf k'},{\bf q};\Omega_{n})
=v^{s}_{\mu}({\bf k},{\bf q})\,v^{s'}_{\nu}({\bf
k'},-{\bf q})\nonumber\\
&\qquad\times[{\cal D}^{+}_{\mu\nu}({\bf q},\Omega_{n})+
(2\delta_{ss'}-1){\cal D}^{-}_{\mu\nu}({\bf q},\Omega_{n})].
\label{eq:effint}
\end{eqnarray}
This describes scattering of CFs from states with spin $s$, $s'$ and momenta
${\bf k}$, and ${\bf k'}$ into states with $({\bf k}+{\bf q})$, $({\bf
  k'}-{\bf q})$ by exchanging a gauge field quantum with momentum ${\bf q}$
and frequency $\Omega_{n}=2\pi nk_{\rm B}T$ ($n$ integer, $k_{\rm B}$
Boltzmann constant, $T$ temperature).

The effective interaction contains the RPA gauge field propagators ${\cal
  D}^{\alpha}_{\mu\nu}({\bf q},\tau)$. In terms of the current-current
correlation functions for free Fermions at zero magnetic field,
$\Pi^{0}_{\nu}\equiv \Pi^{0}_{\nu\nu}({\bf q},\Omega_{n})$ one has
\begin{equation}
({\cal D}^{-1})^{\alpha}_{\mu\nu}({\bf q},\Omega_{n})=
 \left(
\begin{array}{cc}
-\Pi^{0}_{0}&
\qquad\frac{ieq}{\phi_{0}}\\
-\frac{ieq}{\phi_{0}}&\qquad
\frac{q^{2}V(q)}{\phi_{0}^{2}}\delta_{+,\alpha}-\Pi^{0}_{1}
\end{array}
\right).
\end{equation}
It can be shown that the dominant small-momentum small-energy
contributions of the above symmetric and antisymmetric propagators
correspond to $\mu=\nu=1$. \cite{Bonesteel96} For $\Omega_{n}\ll
v_{{\rm F}}q\ll v_{{\rm F}} k_{{\rm
    F}}$, $\Pi^{0}_{0}\simeq e^{2}m^{}/\pi$, $\Pi^{0}_{1}\simeq
-(e^{2}q^{2}/12\pi +2|\Omega_{n}|e^{2}\rho/v_{{\rm F} }q)/m^{}$, such that
\begin{eqnarray}\label{eq:d}
{\cal D}^{+}_{11}({\bf q},\Omega_{n})&\approx&
\frac{-q}{\alpha_{+}(q)\, q^{2}+\alpha_{-}q^3 +\eta\, |\Omega_{n}|}
\nonumber\\
{\cal D}^{-}_{11}({\bf q},\Omega_{n})&\approx&
\frac{-q}{\alpha_{-}\, q^{3}+\eta\, |\Omega_{n}|}
\end{eqnarray}
with the constants $\eta=2e^{2}\rho/m^{}v_{{\rm F}}$, $\alpha_{-}=4\pi/3
m^{}\phi_{0}^{2}$. The function $\alpha_{+}=qV(q)/\phi_{0}^{2}$ depends on the 
nature of the interaction between the electrons. For Coulomb interaction, 
$V(r)=e^2/\epsilon r$, one has $V(q)=2\pi e^2/\epsilon q$. An estimate of the 
magnitude of this energy is given by $E_{\rm C}=e^2/\epsilon l_{\rm B}$. In this 
case, $\alpha_+(q)=const$; for small wave numbers and frequency $\Omega_{n}\to 0$, 
the subleading $\propto q^3$ term in the denominator of ${\cal D}^{+}_{11}({\bf
  q},\Omega_{n})$ can be neglected and the antisymmetric propagator ${\cal
  D}^{-}_{11}({\bf q},\Omega_{n})$ dominates. This can be physically
understood considering
 that the long-range, Coulomb interaction strongly suppresses 
the in-phase density fluctuations
described by $a^+$ in the long wavelength limit.\cite{ketal01}
 
For a short range interaction of the form $V(r)=e^{-r/r_0}e^2/\epsilon r$,
with $r_0\equiv q_{0}^{-1}$ the screening length, $V(q)=2\pi
e^2/\epsilon\sqrt{q^2+q_0^{2}}$.  In order to investigate the influence of the range
of the interaction on the results, we consider below the zero-range limit
$V(q\to 0)$. With this, $\alpha_+(q)\propto q$ and there is no subleading term
in ${\cal D}^{+}_{11}$ of (\ref{eq:d}). 
The in-phase and out-of-phase propagators are of the
same order.

\subsection{The ground state energy.}
The effective interaction (\ref{eq:effint}) 
 turns out to be attractive for Cooper pairs of CFs
(${\bf k}=-{\bf k'}$ and $s=-s'$). This results in the formation of a
condensate of spin singlet Cooper pairs of particles.\cite{metal02}

However, the same interaction provides also the possibility of pairing between
particles with momentum ${\bf k}$ and spin $s$ and holes with ${\bf k}'={\bf k},s'=-s$. The question
arises about which of the two anomalous states is the ground state. In order to
discuss this it is necessary to consider the energies of the two competing
ground states. Below, we introduce two different matrix Green functions ${\cal
  G}$ for the p-p and p-h channels that describe the
properties of the anomalous state they refer to.

The difference in ground state energies per unit area  between the free ($E_0$) and
the interacting ($E$) system described by ${\cal G}$ is obtained introducing a
supplementary coupling constant $\lambda$. Passing to the retarded Green functions 
in the zero temperature limit one has the general expression\cite{agd}
\begin{eqnarray}\label{eq:gs} 
E-E_0&=&
-\int_0^1\frac{{\rm d}\lambda}{\lambda}\int \frac{{\rm d}{\bf 
q}}{(2\pi)^2}\int \frac{{\rm d}\epsilon}{2\pi} 
\Theta(-\epsilon)  \nonumber \\
&&\qquad\,\times{\rm Tr}\, G_0^{-1}({\bf q},\epsilon)
{\,\rm Im}G({\bf q},\epsilon;\lambda).
\end{eqnarray}
The variable $\lambda$ in the retarded Green function $G({\bf
  p},\epsilon;\lambda)$ enters as a switching-on parameter for the effective
interaction $\lambda V^{s,s'}_{\mu\nu}$ and $\Theta(-\epsilon)$ is the Heaviside step 
function.

\subsection{The number of particles.}

At zero temperature the total number of particles is related to the retarded 
Green function $G$ according to\cite{agd}
\begin{equation}\label{eq:number}
N=-\frac{1}{\pi}\int{\rm d}\epsilon \int \frac{{\rm d}{\bf q}}{(2\pi)^2} {\rm 
Tr}\,{\rm Im}\,G({\bf q},\epsilon) \Theta(-\epsilon);
\end{equation}
this implicitly defines the chemical potential $\mu$ in $G$.

In the next section we study the case of Coulomb interaction.  We investigate the
particle-hole condensate while recalling from earlier work the main results for the 
Cooper channel.\cite{metal02,pss,varenna}

%***************************************************************************
%                                                       LONG RANGE INTERACTION
%*****************************************************************************

\section{Long-range interaction}
\label{longrange}

\subsection{The particle-hole channel.}
We calculate in this section the energy gap for the
 particle-hole channel using the Eliashberg technique \cite{Eliash} in
mean field approximation \cite{Khves93}. We introduce a Nambu field
\begin{equation}
{\bf \Phi}({\bf k},\tau)=\left(
\begin{array}{c}
\psi_\uparrow({\bf k},\tau) \\
\psi_\downarrow({\bf k},\tau) \\
\end{array}
\right)
\end{equation}
with $\psi_s({\bf k},\tau)$ the Fermion annihilation operator for spin $s$ and 
momentum ${\bf k}$ at imaginary time $\tau$. It is assumed that terms of the form 
$\langle \psi_\uparrow \psi_\downarrow^\dagger\rangle$,
the so-called anomalous averages that appear in the off-diagonals of the Green
functions ${\cal G} ({\bf k},\tau)=-\langle T_\tau \Phi ({\bf k},\tau){
  \Phi}^\dagger({\bf k},0) \rangle$, are different from zero.  The Green
function ${\cal G}$ is a $2\times 2 $ matrix that obeys the Dyson equation 
\begin{equation}\label{eq:dyson}
{\bf {\cal G}}^{-1}({\bf k},\omega_{n})= {\bf {\cal G}}_{0}^{-1}({\bf
k},\omega_{n}) -{\mathit\Sigma}({\bf k},\omega_{n}). \label{Dyson}
\end{equation}
with ${\cal G}_{0}({\bf k},\omega_{n})= \sigma_0(i\omega_n-k^2/2m+\mu)$ the Green 
function for free Fermions ($\sigma_0$=2$\times$2 identity matrix,  
$\omega_n=(2n+1)\pi k_{\rm B}T$ fermionic frequency). The dominant contribution to 
the Fock self-energy $\mathit\Sigma$ in terms of the effective interaction 
is\cite{Khves93,Nagaosa90}
 \begin{eqnarray} 
\label{eq:self} \mathit\Sigma_{ij}({\bf k},\omega_n)&=& 
k_{\rm B}T\!\!\int\frac{{\rm 
d}{\bf q}}{(2\pi)^2}
\sum_{\Omega_m}{\cal G}_{ij}({\bf k-q},\omega_n-\Omega_m)
           \nonumber \\
&&\qquad\times [ \delta_{ij}   
V_{11}^{s,s} ({\bf k}, {\bf k},{\bf q};\Omega_m) 
 \nonumber \\
&& +(\delta_{ij}-1) V_{11}^{s,-s}({\bf k},-{\bf k},{\bf q};\Omega_m) ] .
\end{eqnarray}
By analytical continuation to real frequencies, $\omega_{n}\to -i\epsilon$, and 
using the spectral representation of the Green function, one obtains implicit 
equations for the retarded self-energies $\Sigma_{11}$ and $\Sigma_{12}$ at 
zero-temperature
\begin{eqnarray}\label{eq:gap}
\Sigma_{11}({\bf k},\epsilon)&=&-\frac{e^2}{2\pi^2m^{2}} \int \!\! 
\frac{{\rm d}{\bf
q}}{(2\pi)^2}
\int_{-\infty}^{+\infty}\!\! {\rm d}\omega {\rm d}\epsilon_1 \nonumber\\
&\times & \!\frac{{\rm Im} \left[D_{11}^+({\bf k-q},\omega)+D_{11}^-({\bf
k-q},\omega)\right]}
{\omega+\epsilon_1-\epsilon-i\delta} \nonumber\\
&\times& \!\frac{({\bf k}\times {\bf q})^2}{\vert {\bf k}- {\bf q} 
\vert^2} ({\rm
sgn}\epsilon_1+{\rm sgn}\omega)  {\rm Im}\,G_{11}({\bf q},\epsilon_1), 
\nonumber\\
\Sigma_{12}({\bf k},\epsilon)&=&-\frac{e^2}{2\pi^2m^{2}} \int \!\! 
\frac{{\rm d}{\bf 
q}}{(2\pi)^2}
\int_{-\infty}^{+\infty}\!\! {\rm d}\omega {\rm d}\epsilon_1 \nonumber\\
&\times & \!\frac{{\rm Im} \left[D_{11}^+({\bf k-q},\omega)-D_{11}^-({\bf
k-q},\omega)\right]}
{\omega+\epsilon_1-\epsilon-i\delta} \nonumber\\
&\times& \!\frac{({\bf k}\times {\bf q})^2}{\vert {\bf k}- {\bf q} \vert^2} ({\rm 
sgn}\epsilon_1+{\rm sgn}\omega) {\rm Im}\,G_{12}({\bf q},\epsilon_1)\nonumber\\
\end{eqnarray}
where $G_{11},G_{12}$ and $D_{11}^\pm$ are the retarded Green functions
continued analytically from ${\cal G}_{11}, {\cal G}_{12}$ and ${\cal
  D}_{11}^\pm$, respectively. The self-energy matrix element
$\mathit\Sigma_{12}=\mathit\Sigma_{21}$ is related to the pairing energy
we are interested in. On the other hand, the diagonal terms of 
$\Sigma_{ij}$ describe usual self-energy corrections; in the 
approximation of constant $\mathit\Sigma_{11}=\mathit\Sigma_{22}$, they only describe corrections 
to the chemical potential.

For analytically estimating  $\Sigma_{ij}({\bf k}, \epsilon)$, we assume that
${k}\approx {k}_{\rm F}$,  and  that $\Sigma$ and consequently 
$G$ do not depend on the direction of the momentum. This corresponds to investigating only
the s-wave pairing, which leads to 
the  isotropy of the ground state.  One gets
\begin{eqnarray}
\label{eq:gapgen}
\Sigma_{ij}(\epsilon)&=& \!\!
\int_{-\infty}^{+\infty}\!\!\!\!\!\! {\rm d}\epsilon_1 \left[ 
O^+(\epsilon,\epsilon_1) + (-1)^{i+j}O^-(\epsilon,\epsilon_1)\right]
\nonumber\\
&&\qquad\qquad\times\int_0^\infty\!\!\!\!
 {\rm d}q\,    
 {\rm Im}\,G_{ij}(q,\epsilon_1)
\end{eqnarray}
where $O^+,O^-$ are the contributions from the symmetric and antisymmetric
gauge field propagators. Their explicit form is given in the Appendix
(\ref{eq:o}).
 
The form of Im$\,G$ is obtained from the Dyson equation (\ref{eq:dyson}) assuming 
negligible imaginary parts of $\Sigma_{ij}$:
\begin{eqnarray}\label{eq:im}
 {\rm Im}G_{11}(q,\epsilon)= -\frac{\pi}{2}
[\delta(\epsilon-\xi_q-\Sigma_{11}(q,\epsilon)
-\Sigma_{12}{}(q,\epsilon))\nonumber \\
+\delta(\epsilon-\xi_q-\Sigma_{11}(q,\epsilon)
+\Sigma_{12}(q,\epsilon)) ] \nonumber\\
 {\rm Im}G_{12}(q,\epsilon)= -\frac{\pi}{2}
[\delta(\epsilon-\xi_q-\Sigma_{11}(q,\epsilon)
-\Sigma_{12}{}(q,\epsilon)) \nonumber \\
-\delta(\epsilon-\xi_q-\Sigma_{11}(q,\epsilon)
+\Sigma_{12}(q,\epsilon)) ]. \nonumber \\
\end{eqnarray}
It implies that the gap $\Delta$ is given
by $\Delta=\Sigma_{12}$. Equations (\ref{eq:gapgen}) have to be solved together with
the constraint (\ref{eq:number}) that describes the dependence of the chemical
potential on the self-energy, assuming $\Sigma_{11},\Sigma_{12} \approx
const$.

As mentioned, $\Sigma_{11}\approx const$ only causes a shift of the chemical
potential, $\mu\rightarrow\overline{\mu}=\mu-\Sigma_{11}$ in both
(\ref{eq:gap}) and ({\ref{eq:number}). The resulting equations are
\begin{eqnarray}
\Delta(\epsilon)&=&\Delta(\epsilon,\overline{\mu},\Delta) \nonumber \\
\overline{\mu}&=&\overline{\mu}(\Delta)
\end{eqnarray}
where 
\begin{equation} \label{eq:mu}
\overline{\mu}=\left\{ \begin{array}{cc}
\mu_0  & \quad {\textrm{for}} \quad \Delta<\overline \mu\\
2\mu_0-\Delta & \quad {\textrm{for}}\quad  \Delta>\overline\mu \\
\end{array}
\right.\,
\end{equation}
is the solution of (\ref{eq:number}) with $\Delta(\epsilon)=\Delta=const$, and 
$\mu_0=k_{\rm F}^2/2m$.

By evaluating separately the contributions of $O^+$ and $O^-$ in
(\ref{eq:gapgen}), one obtains for $\epsilon \to 0$ the self-consistency
condition (cf. ({\ref{eq:deltameno}) and (\ref{eq:deltapiu}))
\begin{equation}\label{eq:selfcons}
\Delta=\Delta^-(\Delta) +\Delta^+(\Delta,E_{\rm C},\Lambda)\,.
\end{equation}
The solution of this is plotted in Fig.~\ref{exc} [$\Lambda$ ultraviolet 
dimensionless cutoff parameter, see (\ref{eq:cutoff})]. This shows that above a 
$\Lambda$-dependent critical value $E_{\rm C}^{\rm cr}$ it is possible to form 
anomalous particle-hole pairs with a gap that is in a very good approximation equal 
to $\mu_0$.
\begin{figure}[htp]
\includegraphics[scale=1]{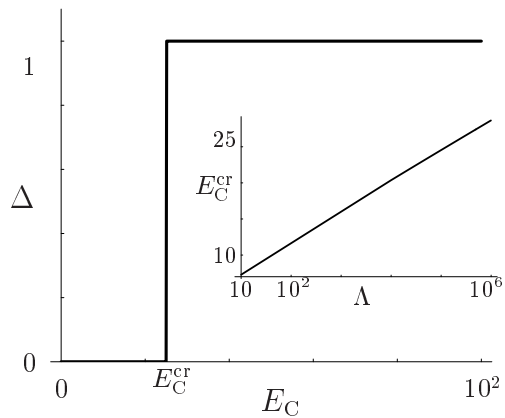}
\caption{ \label{exc} Particle-hole gap $\Delta$ (unit $\mu_{0}$)
  as a function of the Coulomb energy $E_{\rm C}$ (unit $\mu_0/ k_{\rm F}
  l_{B}$) for the cutoff value $\Lambda=10^5$. Inset: dependence of the
  critical value $E_{\rm C}^{\rm cr}$ on $\Lambda$.}
\end{figure}

\subsection{The particle-particle channel.}
Equations (\ref{eq:self}) and (\ref{eq:gapgen}) are written in a form which
also holds for particle-particle pairing. However, in this case $\cal G$ is
the Green function for the Nambu field ${\bf \Phi}^\dagger({\bf
  k},\tau)=\left( \psi^\dagger_\uparrow({\bf k},\tau), \psi_\downarrow({-\bf
    k},\tau) \right)$. This changes $G_{11},G_{12}$ but leaves $O^+,O^-$
invariant. Due to the strong similarities in the formal approaches of the two cases, 
we can simply use the previous results. \cite{metal02,varenna} The pair breaking gap 
is given by
\begin{equation}
\label{eq:deltasc} 
\Delta(\epsilon)=\frac{\epsilon\Sigma_{12}(\epsilon)}{\epsilon-
\Sigma_{11}(\epsilon)}.
\end{equation}
The equation to be solved for $\Delta(0)=\Delta$ is
\begin{equation}
\label{eq:scdelta}
 1=\frac{C_-}{\Delta^{1/3}}-\frac{1}{2\pi E_{\rm C}} 
\log^2\!\frac{\Delta}{\Lambda E_{\rm C}}=
f(\Delta,E_{\rm C},\Lambda) 
\end{equation}
with $\Delta$ in units of $\mu_0$, $E_{\rm C}$ in units of $\mu_0/k_{\rm
  F}l_{\rm B}$, $C_-\approx1.4$ and $\Lambda$ is a cutoff parameter. 
The solutions are shown in Fig.~\ref{sc}.
\begin{figure}[htb]
\includegraphics[scale=1]{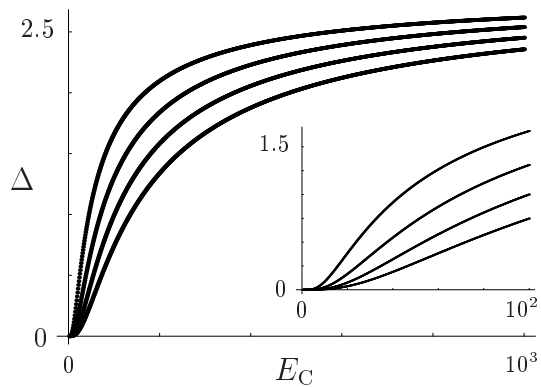}
\caption{\label{sc} Particle-particle gap $\Delta$ 
  as a function of the Coulomb energy $E_{\rm C}$ for different values of the
  cutoff $\Lambda$ ($=10^3, 10^4, 10^5, 10^6$, top to bottom, units as in
  Fig.~\ref{exc}). Inset: small-$E_{\rm C}$ behavior. }
\end{figure}

Here, a solution for $\Delta$ is found for any value of $E_{\rm C}$: nevertheless, 
values of $\Delta>1$ are  not to be accepted because they are outside the range of 
validity of the assumptions in the calculations.\cite{metal02,varenna}
 For large 
$E_{\rm C}$, $\Delta$ is nearly independent of the Coulomb energy and is determined 
only by the $O^-$ contribution, $\Delta=C_-^3$: this is consistent with the 
statements in Sec.(\ref{section:effint}) about the gauge field propagators. In fact, 
a strong Coulomb interaction  quenches the in-phase density fluctuations described 
by $a^+$ and makes the $D^-$ contribution even more dominant for $q\to 0$.

\subsection{The phase diagram}
In order to compare the two pair states one has to compare the gains in their ground 
state energies with respect to the non-paired state.  Equation (\ref{eq:gs}) gives 
the energy difference between the non-interacting system and the interacting system 
described by $G$. As we are interested in the difference between the energies of the 
anomalous state $E_{}$ and the normal interacting system $E_{\rm n}$, we write
\[ E_{}-E_{\rm n}=(E_{}-E_0)-(E_{\rm n}-E_0) \]
and perform the calculations in (\ref{eq:gs}) twice, first with the full $G$
and then with $G$ for $\Sigma_{12}\to 0$.

\subsubsection{Particle-hole ground state energy.}\label{par:phgse}
In this case we use the expressions for the imaginary parts in (\ref{eq:im}) and 
$G_{011}^{-1}(q,\epsilon)=G_{022}^{-1}(q,\epsilon)= \epsilon-q^2/2m^{}+\mu$. 
Performing the $q-$ and $\epsilon-$ integrations we find
\[ E_{}-E_{\rm n}=-\frac{m^{}}{2\pi}\int_0^1 \frac{{\rm d}\lambda}
{\lambda}\Delta^2_\lambda. \] The gap $\Delta_\lambda$ is the solution of
(\ref{eq:selfcons}) with an effective interaction $\lambda V^{s,s'}_{\mu\nu}$.
It can be shown that
\begin{equation}
\label{eq:la} \Delta_\lambda=\lambda\left[\Delta^+(\Delta_\lambda) 
+\Delta^-(\Delta_\lambda,E_{\rm C},\Lambda)\right].
\end{equation} 
From the numerical analysis of this, we know that there is a critical value
$\lambda^{\rm cr}(E_{\rm C},\Lambda)$ below which $\Delta_\lambda=0$, otherwise
$\Delta_\lambda\approx\mu_0$. 

To obtain the dependence of the critical parameter on $E_{\rm C}$ and
$\Lambda$, we solve (\ref{eq:la})  for
$\lambda^{\rm cr}$ in the limit $\Delta_\lambda\to \mu_0$. This gives
 \begin{equation}
E_{}-E_{\rm n}\approx\frac{m^{}}{2\pi} 
\Delta^2 \log \lambda^{\rm cr}. 
\end{equation}
The energy gain per particle  is then
\begin{equation} \label{eq:gainexc}
\Delta E = \frac{\vert E-E_{n}\vert }{\rho}=\frac{\mu_0}{2} |  \log \lambda^{\rm cr} |. 
\end{equation} 
 This is plotted in Fig.~\ref{10a5} as a function of $E_{\rm C}$(thin line). 
Since $\Delta=0$ for $E_{\rm
  C}<E_{\rm C}^{\rm cr}(\Lambda)$, $\Delta E=0$ in that region.

\subsubsection{Particle-particle ground state energy.}
\begin{figure}[htp]
\includegraphics[scale=1]{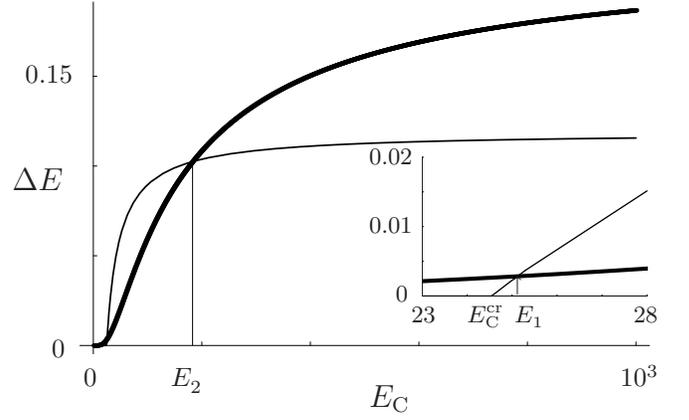}
\caption{ \label{10a5} The energy gain $\Delta E$ per particle  
  as a function of $E_{\rm C}$ for $\Lambda=10^5$ (units as in
  Fig.~\ref{exc}). Thick curve: p-p state; thin curve: p-h state. Inset:
  region near $E_{\rm C}^{\rm cr}$ ($E_{\rm C}^{\rm cr}$ critical energy for
  p-h pair formation, $E_{1,2}^{}$ energies at which the curves
  intersect (see text).}
\end{figure}

In order to use (\ref{eq:gs}) on the p-p channel the components of $G_0$
and $G$ are required,
\begin{equation}
G_{011}^{-1}(q,\epsilon)=\epsilon-\xi_q \, , \quad 
G_{022}^{-1}(q,\epsilon)=\epsilon+\xi_q 
\end{equation}
and 
\begin{eqnarray}
\label{eq:imgsc}
{\rm Im}G_{i}(q,\epsilon)&=&-\pi\,{\rm sgn} 
(\epsilon-\Sigma_{11}(\epsilon))\times\nonumber\\
&\times&\frac{N_i}{2\Omega}\left[\delta(\xi_q-\Omega)+
\delta(\xi_q+\Omega)\right],
\end{eqnarray}
with index 
$i$ denoting ${11}$ or ${12}$ and 
\begin{eqnarray*}
N_{11}&=&\epsilon-\Sigma_{11}(\epsilon)+\xi_q \nonumber\\
N_{12}&=&\Sigma_{12}(\epsilon). 
\end{eqnarray*}
We have here also defined $\Omega=[(\epsilon-\Sigma_{11}(\epsilon))^2-
\Sigma_{12}(\epsilon)^2]^{1/2}$, $\xi_q=q^2/2m^{}-\mu$ and the
off-diagonal component $G_{12}$ has been introduced for later convenience.
Since we have neglected the even part of $\Sigma_{11}$, Im$\,G_{11}=$
Im$\,G_{22}$ which also explains why the chemical potential is not modified in
the full Green function. Using (\ref{eq:deltasc}) and the parity properties of
$\Sigma_{11}, \Sigma_{12}$ one finds
\begin{equation}\label{eq:gg}
E_{}-E_0=-\frac{m^{}}{2\pi} 
\int_0^1\frac{{\rm d}\lambda}{\lambda}
\int_{\Delta_\lambda}^\infty  \!\!\!{\rm d}\epsilon 
\frac{\epsilon \Sigma_{11}(\epsilon)+ \Delta_\lambda 
\Sigma_{12}(\epsilon)}
{\sqrt{\epsilon^2-\Delta_\lambda^2}} .
\end{equation}
Subtracting the same quantity with $\Delta_\lambda\to0$ gives
\begin{equation}
E_{}-E_{\rm n} \approx \!
-\frac{m^{}}{2\pi} \!\!\int_0^1\!\frac{{\rm d}\lambda}{\lambda}
\!\left[\int_{\Delta_\lambda}^\infty  \!\!\!\!{\rm d}\epsilon \frac{\Delta_\lambda^2}
{\epsilon} \!-\!\!\!\int_0^{\Delta_\lambda} \!\!\!\! {\rm d}\epsilon 
\Sigma_{11}^{\Delta=0}(\epsilon) \right],
\end{equation}
by expanding the integrand in (\ref{eq:gg}) for $\epsilon\gg\Delta_\lambda$ and
assuming $\Sigma_{11}(\epsilon\gg\Delta_\lambda)\approx
\Sigma_{11}^{\Delta=0}(\epsilon)$. The first part should be integrated with a cutoff 
$\Lambda_{\rm C}$ and would give a logarithmic contribution 
$\propto\Delta_\lambda^2\log \Delta_\lambda/\Lambda_{\rm C}$.  The most important 
contribution\cite{b99,ul94}  comes from the second integral that can be evaluated 
explicitly to the same accuracy  taking into account in  (\ref{eq:gapgen}) only 
$O^-$ in the limit $\Delta\to 0$ 
\begin{equation}\label{eq:scgain}
E_{}-E_{\rm n} \approx \frac{ m^{}}{2\pi} \int_0^1\frac{{\rm 
d}\lambda}{\lambda} \int_0^{\Delta_\lambda} \!\!\!\!\!\! {\rm d}\epsilon  
\Sigma_{11}^-(\epsilon). 
\end{equation}
 In the same
limit\cite{metal02}
\begin{eqnarray} 
\label{eq:intimg}
 \int {\rm d}q {\rm Im}G_{11}(q,\epsilon_1) &=& -\frac{\pi m^{}}{ k_{\rm F}}
\end{eqnarray}
and 
\begin{equation}
 \int_0^{\Delta_\lambda} \!\!\!\!\!\! {\rm 
d}\epsilon \Sigma_{11}^-(\epsilon) = A^- \frac{27\pi m^{}}{5k_{\rm F}} \Delta_\lambda^{5/3}=
B^- \mu_0^{1/3} \Delta_\lambda^{5/3}, 
\end{equation}
using the notations of (\ref{eq:const}) for the value of the constant $A^-$ and 
implicitly defining the numerical constant $B^-$.

The energy \emph{gain} per particle is then 
\begin{equation}
\Delta E=\frac{\mu_0}{2} \, \vert B^- \vert \int_0^1 \frac{{\rm 
d}\lambda}{\lambda}   
 \left(\frac{\Delta_\lambda}{\mu_0}\right)^{5/3}.
\end{equation}
 The final step has been then performed
numerically with the self-consistency equation (\ref{eq:scdelta}) 
modified according to $1=\lambda f(\Delta_\lambda,E_{\rm C},\Lambda)$. 
The results are shown in Fig.~\ref{10a5}
(bold curve).

\subsubsection{Phase diagram.}
Figure \ref{10a5} shows that the curves for $\Delta E$ corresponding to the
two models intersect at certain energies $E^{}_{1}(\Lambda)$ and $E^{}_{2}(\Lambda)$. 
These energies separate the regions of stability of
excitonic and Cooper pair phases.
 
It is useful to recall the validity of the assumptions made in the calculations. 
First, the validity of a mean-field treatment of the interaction has to be 
addressed. It has been shown\cite{SternHalp95} that in the normal state the dominant 
contribution of the gauge field propagator is not expected to be renormalized by 
vertex corrections. It is not clear whether or not this approximation still holds in 
the anomalous states.\cite{Khves93} For approaching the paired state from the normal 
state, we believe that neglecting vertex corrections, and using the bare vertices 
(\ref{eq:vertex}) in (\ref{eq:effint}) is at least a reasonable starting point.  
Second, earlier calculations\cite{metal02} show that the particle-particle energy 
gap survives the linearization of the dispersion law of the fermions around the 
Fermi level. However, for the particle-hole channel it is necessary to keep a higher 
accuracy and take the full quadratic dependence of $\xi_q$ on $q$ into account (cf. 
(\ref{eq:intimgexc})).

For energies $E_{1}^{}<E_{\rm C}<E_{2}^{}$ it is more favorable to form a p-h state, while for
$E_{\rm C}<E_{1}^{}$ and $E_{\rm C}>E_{2}^{}$ the formation of a
p-p state is energetically favorable. The dependence on the value of the
cutoff of the threshold energies is shown in Fig.~\ref{diagram}. Light grey
regions corresponds to the p-p states. The dashed line corresponds to the
values of $E_{\rm C}$ such that the p-p pair breaking gap equals $\mu_{0}$.
Due to the approximations used in the calculations of the p-p gap,
 only the part of the graph to the left of this
curve can be expected to describe correctly the system.

\begin{figure}[htp]
\includegraphics[scale=1]{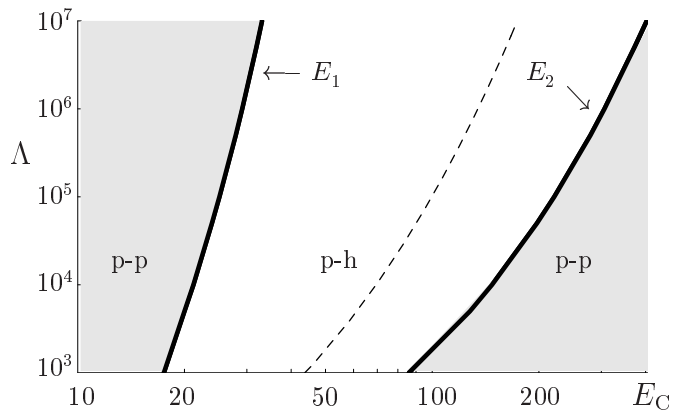}
\caption{ \label{diagram} The phase diagram for the ground state 
  in the plane of the Coulomb energy $E_{\rm C}$ and the cutoff parameter
  $\Lambda$. Light grey region: the p-p state; white region: p-h state.
  Dashed: $E_{\rm C}(\Delta=\mu_{0})$ for the p-p state.}
\end{figure}

%*****************************************************************************
%                                                      SHORT RANGE INTERACTION
%*****************************************************************************

\section{Short-range interaction}
\label{section:short}
\label{shortrange}

The possibility of a crossover between an excitonic and a superconducting CF
state has been demonstrated for long range Coulomb interaction. In this
section we want to investigate whether or not this is a generic feature of any
interaction. We consider an interaction potential of the form introduced in
Section (\ref{sec:effint}),
\[ V(q)= \frac{2\pi 
  e^2}{\epsilon\sqrt{q^2+1/r_0^2}}.\] It has been pointed out above that in
this case one must treat the in-phase and out-of-phase gauge field fluctuations 
$D^+$ and $D^-$ on the same footing. By defining $V_0=V(q\to 0)=2\pi e^2 
r_0/\epsilon$ and $\alpha_+'=\alpha_+/q=V_0/\phi_0^2$ one obtains for the bosonic 
propagators
\begin{eqnarray}\label{eq:dpiumenosr}
{\cal D}^{+}_{11}({\bf q},\Omega_{n})&\approx&
\frac{-q}{(\alpha_+'+\alpha_-)\, q^{3} +\eta\, |\Omega_{n}|} \nonumber\\
{\cal D}^{-}_{11}({\bf q},\Omega_{n})&\approx&
\frac{-q}{\alpha_{-}\, q^{3}+\eta\, |\Omega_{n}|}\,.
\end{eqnarray}
Thus one proceeds along the line of the calculations done for long-range
interaction for the case of $D^-$.

\subsection{The particle-hole state}

To find the gap we have to solve (\ref{eq:gapgen}) for $\Sigma_{12}$, but now
$O^\pm$ have to be calculated according to (\ref{eq:opiumeno}) with ${\cal
  D}^\pm$ in  (\ref{eq:dpiumenosr}). For the real parts one gets
\begin{eqnarray}
O^\pm(\epsilon,\epsilon_1)&=& A^\pm_{\rm sr}\, \frac{ 1+ 3 \, 
{\rm sgn} \epsilon_1 \, {\rm 
sgn}(\epsilon_1-\epsilon) } {(\epsilon-\epsilon_1)^{1/3}}
\end{eqnarray}
with
\begin{eqnarray*}
A^-_{\rm sr}&=&C\frac{\pi^2}{9}\frac{1}{\alpha_-^{2/3}\eta^{1/3}} 
\nonumber \\
A^+_{\rm sr}&=&C\frac{\pi^2}{9}\frac{1}{(\alpha_+'
+\alpha_-)^{2/3}\eta^{1/3}}=A^-_{\rm sr}
\left(1+\frac{\alpha_+'}{\alpha_-}\right)^{-2/3}
\end{eqnarray*}
and $C$ in (\ref{eq:ccc}).
\begin{figure}[htp]
\includegraphics[scale=1]{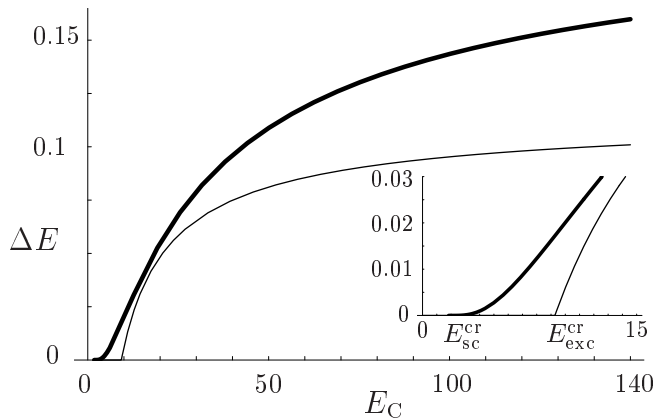}
\caption{ \label{fig:short} The energy gain $\Delta E$ per 
  particle (units $\mu_0$) for short range interaction as a function of
  $E_{\rm C}$ (units $4\mu_0 l_{B}/3r_0(k_{\rm F}l_{B})^2$); p-p state
  (bold curve), p-h state (thin curve). Inset: the region near $E_{\rm
    sc}^{\rm cr}$and $E_{\rm exc}^{\rm cr}$.}
\end{figure}

The integrals which one has to evaluate have the same structure as for $\Delta^-$ 
[cf. (\ref{eq:menodelta})] in the long-range case. The result is very similar and 
the self-consistency equation is (for $\Delta<\mu_0$)
\begin{equation}
\label{eq:srdelta}
 \Delta=K^- A\,\Gamma^<\Big[(\mu_0+\Delta)^{1/6}-(\mu_0-\Delta)^{1/6}\Big] 
\end{equation}
with the prefactor 
\[A= 1-\left(1+\frac{\alpha_+'}{\alpha_-}\right)^{-2/3}\]
and the constants $K^-, \Gamma^<$ of (\ref{eq:come}, \ref{eq:come2}). It can be seen 
by numerical evaluation that this equation has always a solution ($\approx \mu_{0}$) 
if
\[ 9.3\approx r_{\rm cr}<r=
\frac{\alpha_+'}{\alpha_-};\]
since
\begin{equation}\label{eq:r}
r=(k_{\rm F}l_{B})^2 \frac{3}{4} \frac{E_{\rm
    C}}{\mu_0}\frac{r_0}{l_{B}},\end{equation} 
from the last
inequality we define a critical value $E_{\rm exc}^{\rm cr}$ for the Coulomb energy
$E_{\rm C}$ below which $\Delta=0$ 
as in the long-range limit (cf. Fig. \ref{diagram},\,\ref{fig:short}).

The ground state energy gain per particle can be estimated along the same line of (\ref{par:phgse}):
the result is as in
(\ref{eq:gainexc})
\begin{equation} \label{eq:gainexcsr}
\Delta E =\frac{\mu_0}{2} |  \log \lambda^{\rm cr}_{\rm sr} | 
\end{equation} 
 but with a critical parameter $\lambda_{\rm
  sr}^{\rm cr}$ obtained from the solution of (\ref{eq:la}) with the 
short-range form of $\Delta^+,\Delta^-$: 
\begin{equation}
\lambda^{\rm cr}_{\rm sr}\approx\frac{0.79}{A}.
\end{equation}

\subsection{The particle-particle state}

It has been shown previously\cite{varenna} that by substituting the appropriate 
Green functions (\ref{eq:imgsc}), the two equations (\ref{eq:gapgen}) can be 
combined according to (\ref{eq:deltasc}) to yield a self-consistency equation for 
$\Delta$, similar to (\ref{eq:scdelta}) but without the $\log^2$ term due to the 
Coulomb interaction. Combining these results one finds
\begin{equation}
\label{eq:scdsr} 1=C_- \left(\frac{\Delta}{\mu_0}\right)^{-1/3} \left[ 1-2 \left(1+ 
\frac{\alpha_+'}{\alpha_-}\right)^{-2/3} \right],
\end{equation}
where 
\[ C_-=-\frac{2m\pi}{k_{\rm F}}\frac{8}{3}A^-_{\rm sr} 
\frac{3\sqrt{\pi}\Gamma(7/6)}{\Gamma(2/3)} \,\mu_0^{-1/3}\approx 1.4 \] is the same 
as in (\ref{eq:scdelta}) and the prefactor reflects the competition of the $O^\pm$ 
contributions.  A critical value for the ratio $r$ exists also in this case due to 
the requirement that $\Delta>0$; one has $r>2^{3/2}-1$. This defines a critical 
value $E_{\rm sc}^{\rm cr}$ according to (\ref{eq:r}); for $E_{\rm C}<E_{\rm 
sc}^{\rm cr}$ the gap equation has no solutions.
 Otherwise, the gap is an
increasing function of $r$ starting from $\Delta=0$ and reaching $\Delta=1$
for $r\approx 16.8$. Equation (\ref{eq:scgain}) and the following one provide
the estimate for the ground state energy difference,
\begin{equation}
E_{}-E_{\rm n}=B^- \mu_0^{1/3}\frac{m}{2\pi} \left(1+(1+r)^{-2/3}\right) \int_0^1 
\frac{{\rm d}\lambda}{\lambda} \Delta_{\lambda}^{5/3} 
\end{equation}
where $B^-\approx -0.39$ is the same as in the Coulomb case since the antisymmetric 
propagator is not affected by the range of the interaction. The index $\lambda$ in 
$\Delta_\lambda$ has the usual meaning: equation (\ref{eq:scdsr}) with a factor 
$\lambda$ added to the right hand side can be used to obtain explicitly 
$\Delta_\lambda$. The energy gain per particle  is
\begin{equation}
\Delta E=\vert B^-\vert  \frac{\mu_0} {2}\frac{C_-^5}{5}\left[1+(1+r)^{ - 
\frac{2}{3}}\right] \left[1-2(1+r)^{-\frac{2}{3}}\right]^5.
\end{equation}
The result is shown in Fig.~\ref{fig:short} as a function
of $E_{\rm C}$ (bold curve).
Within the present approximations, the two curves for the p-p state and the
p-h state do not intersect. The energy gain per particle for the p-p state is
always larger than for the p-h state if the interaction is short ranged. The
excitonic state is always suppressed in favour of the Cooper pair-like state.

%*****************************************************************************
%                                                                   CONCLUSION
%*****************************************************************************

\section{Conclusions}
\label{conclusion} We have investigated in this paper whether or not the residual 
interaction due to fluctuations of the CS gauge field between CFs with spin at 
filling factor 1/2 can lead to the formation of new collective ground state. We have 
assumed that fluxes and Fermions corresponding to the {\em same} spins are coupled 
via the CS transformation. We take into account the renormalization of the 
propagator of the gauge field due to the coupling to the Fermions. The dominant 
effective interaction between the CFs is then of second order in the gauge 
field-electron vertex and we have found that it can be attractive between CFs with 
opposite spins. Thus, the formation of pairs is possible, which we have investigated 
in the spin singlet, s-wave channel.

We have estimated both the pair-breaking gaps and the ground state energies of 
particle-particle and particle-hole channels for long-range Coulomb and finite range 
interactions. We find that in the former case both, the particle-particle as well as 
the particle-hole state can be stable depending on the strength of the interaction. 
The particle-particle state is stable if the Coulomb energy is smaller than a 
certain threshold energy (which depends on a cutoff parameter). For higher Coulomb 
energy, the excitonic state is favored. If the interaction is screened, symmetric 
and antisymmetric density fluctuations, as described by $D^+$ and $D^-$, become 
comparable and the particle-particle state is always more stable than the 
particle-hole state.

The formation of these states has been shown to be possible if the spin $\uparrow$ 
and the spin $\downarrow$ Landau levels are degenerate and both of them exactly at 
filling factor 1/2. One can suspect that the results remain valid also if these 
conditions are not exactly fulfilled. If the two Landau levels are not degenerate 
the new ground state will form as long as the energy separation between the two 
levels is smaller than the gain in the ground state energy. Assume the level with 
spin $\uparrow$ to be energetically lower. Then, the ground state without 
interaction corresponds to this level completely filled. 
With interaction (i.e. with gauge field fluctuations), however, the instabilities 
discussed in the present paper would be
 present and half of the electrons would occupy the (energetically higher) level with spin 
$\downarrow$ such that the energetically more favorable collective ground state can be 
achieved by forming spin singlet particle-particle or particle-hole pairs. The 
situation in which each of the two levels is exactly at $\nu=1/2$ is then the ground 
state since any deviation from this occupation would yield a higher energy.
This mechanism\cite{varenna}
would be relevant in the interpretation of the intermediate plateaus in the
 optical measurements of spin polarization\cite{Kukushkin} for both the p-p and 
the p-h pairings; thus further experimental investigation would be necessary to 
test the actual interplay between the two proposed phases.

%*****************************************************************************
%                                                                     APPENDIX
%*****************************************************************************

\appendix*  
\section{Details of calculations}
 \label{sec:appendix}
 In the evaluation of the integrals we use similar results for the Cooper
 channel.\cite{pss} In order to perform the ${\bf q}$-integration in Eq.
 (\ref{eq:gap}) and get the form in Eq.(\ref{eq:gapgen}), we rewrite the
 expression for the vertices with $p=\vert{\bf k-q}\vert$,
\begin{equation}
\frac{({\bf k}\times {\bf q})^2}{\vert {\bf k}- {\bf q}
\vert^2}=\frac{k^2q^2}{p^2}\sin^2\theta
\end{equation}
where $\theta$ is the angle between ${\bf k}$ and ${\bf q}$. Aligning the
$q_x$ axis along the $\hat{k}$ direction, the measure is changed
\begin{equation}
  \label{eq:help}
\int_0^\infty \!\!q{\rm d}q\int_0^{2\pi}\!\!{\rm d}\theta 
= 2 \int_0^\infty \!\!{\rm d}q
\int_{\vert k-q\vert}^{k+q} \!\!\frac{ p{\rm d}p}{k\sin\theta}
  \end{equation}
with
\begin{equation}
\sin\theta=\left[1-\Big(\frac{k^2+q^2-p^2}{2kq}\Big)^2\right]^{1/2}\,.
\end{equation}
If we assume for the external momentum $k\approx k_{\rm F}$ and consider only the 
dominant contribution with $q\sim k_{\rm F}$, we get for 
$\Sigma_{ij}(\epsilon)\approx\Sigma_{ij}(k_{\rm F},\epsilon)$ ($i=1,j=1,2$)
\begin{eqnarray}
\Sigma_{ij}(\epsilon)&=& C \int_0^\infty \!\!  {\rm d} q \int_0^{2k_{\rm F}} 
\!\!\!\!{\rm d}p \sqrt{1-\frac{p^2}{4k_{\rm F}^2}}\times  \nonumber\\
&&\times\int_{-\infty}^{+\infty}\!\!\!\! 
{\rm d}\omega {\rm d}\epsilon_1 ({\rm sgn}\epsilon_1+{\rm sgn}\omega)  
{\rm Im}G_{ij}(q,\epsilon_1)\times  \nonumber\\
&&\times\frac{{\rm Im} \left[D_{11}^+({p}, \omega)+(-1)^{i+j}
D_{11}^-({p},\omega)\right]} 
{\omega+\epsilon_1-\epsilon-i\delta} 
\end{eqnarray}
with the constant
\begin{equation}\label{eq:ccc}
C=-\frac{e^2 k_{\rm F}^2}{4\pi^4 m^2}. \\
\end{equation}
In order to obtain (\ref{eq:gapgen}) we then have
\begin{eqnarray} 
\label{eq:o}
O^{\pm}(\epsilon,\epsilon_1)&=&C \int{\rm d}p \int{\rm d}\omega 
({\rm sgn}\epsilon_1+{\rm sgn}\omega) \nonumber\\
&&\qquad\times \frac{ {\rm Im} D^{\pm}_{11}(p,\omega)} 
{\omega+\epsilon_1-\epsilon-i\delta} .
\end{eqnarray}
Assuming $p\ll k_{\rm F}$ the $p$-integral can be performed,
\begin{eqnarray}
{\rm Im} D_{11}^{\pm}(p,\omega)= \frac{-\eta\omega p}{\alpha_{\pm}^2 p^{(5\mp 
1)}+\eta^2\omega^2 } 
\end{eqnarray}
and
\begin{eqnarray} 
\label{eq:opiumeno}
  \int_0^\infty {\rm d}p \, {\rm Im} 
D_{11}^{+}(p,\omega)&=&-\frac{\pi}{4\alpha_+} {\rm sgn} \omega \,,
\nonumber \\
\int_0^\infty {\rm d}p \, {\rm Im} D_{11}^{-}(p,\omega)& =&-\frac{\pi}{3\sqrt{3}} 
\frac{1}{\alpha_-^{2/3}\eta^{1/3}\omega^{1/3}}.\nonumber\\
\end{eqnarray}
Now the energy integrations have to be performed as principal value integrals. We 
have
\begin{widetext}
\begin{eqnarray}
O^+(\epsilon,\epsilon_1)&=& A^+ \left[\log  \frac{|\Lambda_{\rm C} 
+\epsilon_1-\epsilon|}{|\Lambda_{\rm C}-\epsilon_1+\epsilon|} + {\rm 
sgn}\epsilon_1\log  \frac{|\Lambda_{\rm 
C}^2-(\epsilon_1-\epsilon)^2|}{|(\epsilon_1-\epsilon)^2|} +i \pi (1-{\rm 
sgn}\epsilon_1 \,{\rm sgn}(\epsilon_1-\epsilon)) \right], \nonumber\\
O^-(\epsilon,\epsilon_1)&=& A^-\left[ \frac{ 1+ 3 \, {\rm sgn} 
\epsilon_1 \, {\rm 
sgn}(\epsilon_1-\epsilon) } {(\epsilon-\epsilon_1)^{1/3}} -
i \sqrt{3}\frac{{\rm 
sgn}\epsilon_1+ {\rm sgn}(\epsilon-\epsilon_1)} 
{(\epsilon-\epsilon_1)^{1/3}}\right],
\end{eqnarray}
\end{widetext}
with the constants
\begin{eqnarray}\label{eq:const}
A^+&= &-C \frac{\pi}{4\alpha_+} \nonumber\\
A^-&= & C \frac{\pi^2}{9} \frac{1}{\alpha_-^{2/3}\eta^{1/3}}
\end{eqnarray}
and a cutoff $\Lambda_{\rm C}$ that must be introduced to evaluate $O^+$. A
physically meaningful value for this can be estimated by considering in more
detail the integral
\[ \int_0^{2k_{\rm F}} {\rm Im}D^+_{11}(p,\omega)=
-\frac{1}{2\alpha_+}\left(\frac{\pi}{2}- \arctan\frac{\eta\omega}{4 k_{\rm
    F}^2\alpha_+}\right). \] 
This vanishes for $\omega\to\infty$. The scale for the
vanishing of the integral can be obtained by considering the argument of the
$\arctan$
\begin{equation} \label{eq:cutoff}
\frac{\eta\omega}{4 k_{\rm F}^2\alpha_+}=
\frac{\omega}{E_{\rm C}}\frac{1}{2k_{\rm F} l_{\rm B}}
\end{equation}
where $E_{\rm C}=e^2/\epsilon l_{\rm B}$. From this, it is reasonable to
choose as the cutoff $\Lambda_{\rm C}=\Lambda k_{\rm F}l_{\rm B}E_{\rm C}$,
where $\Lambda$ represents the numerical value of the cutoff.

Next step is to consider the contribution from the fermionic Green function:
the $q$-integrals for the diagonal part, Im$G_{11}$, and the off-diagonal
part, Im$G_{12}$, yield
\begin{eqnarray}\label{eq:intimgexc}
\Theta^{ij}(\epsilon)&\equiv&\int \!\!
{\rm d}q {\rm Im}G_{ij}(q,\epsilon)=\nonumber\\
&=&-\frac{\pi}{2} \sqrt{\frac{m}{2}} 
\left[\frac{\theta(\epsilon+\overline\mu-\Delta)}
{\sqrt{\epsilon+\overline\mu-\Delta}}\right.+\nonumber\\
&&+(-1)^{i+j}\left.
\frac{\theta(\epsilon+\overline\mu+\Delta)}
{\sqrt{\epsilon+\overline\mu+\Delta}}\right]
\end{eqnarray}
This gives finally for the self-energies
\begin{equation}
\Sigma_{ij}(\epsilon)=\int_{-\infty}^{+\infty} {\rm d}\epsilon_1 
\Theta^{ij}(\epsilon_1) \left[O^+(\epsilon,\epsilon_1)+(-1)^{i+j}
O^-(\epsilon,\epsilon_1)\right]. 
\end{equation}

We now concentrate on the $\epsilon_1$-integral for $\Delta=\Sigma_{12}$ in
the limit $\epsilon \to 0$. We first realize that the imaginary parts of $O^+$
and $O^-$ do not contribute. The main contribution comes
from Re$\,O^-$ and, with the notation of (\ref{eq:selfcons}),

\begin{equation} 
\label{eq:deltameno} 
\Delta^-= -\int{\rm d}\epsilon_1 
\Theta^{12}(\epsilon_1)O^-(\epsilon\to 0,\epsilon_1). \nonumber
\end{equation}
The latter integral can be solved in the two regimes 
$\Delta\lessgtr\overline\mu$. 
One finds
\begin{equation}
\label{eq:menodelta}
\begin{array}{clc}
\Delta^-=&  K^- \Gamma^< \left[ 
(\overline\mu+\Delta)^{1/6}-(\overline\mu-\Delta)^{1/6}\right] \,  
&(\Delta<\overline\mu)\nonumber  \\
&&\nonumber  \\
\Delta^-=& K^-\Big[\Gamma^> (\Delta-\overline\mu)^{1/6}+ 
\Gamma^<(\Delta+\overline\mu)^{1/6} \Big]
 \,  &
\, (\Delta>\overline\mu) 
\end{array}
\end{equation}
with the constants $\Gamma^>,\Gamma^<$ defined in terms of Euler gamma
function $\Gamma$
\begin{eqnarray}\label{eq:come}
 \Gamma^<&=&\Gamma\left(\frac{2}{3}\right)\left(\frac{\sqrt{\pi}} 
{\Gamma(7/6)}-\frac{\Gamma(-1/6)}
{\sqrt{\pi}}\right) \approx 7.76\nonumber \\
&&\nonumber \\
\Gamma^>&=&\frac{\sqrt{\pi}\Gamma(-1/6)}{\Gamma(1/3)} \approx -4.48
\end{eqnarray}
and
\begin{eqnarray}\label{eq:come2} 
 K^-&=& -4\frac{\pi}{2}\sqrt{\frac{m}{2}}A^- \approx 0.15 \mu_0^{5/6} .
\end{eqnarray}
These must be combined with the corresponding relations for
$\overline\mu=\overline\mu(\Delta)$ of (\ref{eq:mu}) to obtain a
self-consistency equation for $\Delta$. Neglecting for the moment the
contribution of $O^+$, we have for $\Delta<1$
\[ \Delta=1.13 \left( (1+\Delta)^{1/6}-(1-\Delta)^{1/6}\right) \]
where $\Delta$ is expressed in units of $\mu_0$. The solution to this
equation is indeed very close to $\mu_0$ itself.

The inclusion of $O^+$ implies the solution of a more complicated integral
\begin{equation} 
\label{eq:deltapiu} 
\Delta^+= \int{\rm d}\epsilon_1 
\Theta^{12}(\epsilon_1) O^+(\epsilon\to 0,\epsilon_1). \nonumber
\end{equation}
It can be solved analytically and it is possible to show that it only shifts
the solution even closer to $\mu_0$. The most important effect of considering
the $O^+$ integral is, however, that it introduces a new energy scale $E_{\rm
  C}$ and a cutoff parameter $\Lambda$. The value of the gap is largely
independent from $E_{\rm C}$, but depending on the cutoff there exists a
critical value $E_{\rm C}^{\rm cr}(\Lambda)$ of the Coulomb energy below which
there are no solutions to the equation $\Delta=\Delta^++\Delta^-$ (see Fig.
~\ref{exc}).

\begin{acknowledgments}
  {\bf Acknowledgments:} We thank Klaus von Klitzing, Rolf Haug, Eros Mariani and
 Franco Napoli for helpful
  and illuminating discussions. Financial support by the European Union
  via 
HPRN-CT2000-0144, from the DFG via
  Special Research Programme ''Quantum Hall Systems'' and from the Italian
  MURST PRIN02 is gratefully acknowledged.
\end{acknowledgments}

\end{document}